\newcommand{\eg}{\emph{e.g.}, }
\newcommand{\ie}{\emph{i.e.}, }
\newcommand{\cf}{\emph{cf.}\ }

\documentclass{article}
\usepackage{graphicx,psfig,epsfig}
\def\title{\begin{center}\Large\bf}
\def\author(s){\vspace{0.3cm}\large\rm}
\def\text{\end{center}}

\begin{document}

\title
Dynamics of triaxial galaxies with a central density cusp

\author(s)
Christos Siopis$^{\rm 1,2}$ and Henry E.\ Kandrup$^{\rm 1,3,4}$

$^{\rm 1}${\it Department of Astronomy, University of Florida, Gainesville, FL 32611, USA \\}
$^{\rm 2}${\it Department of Astronomy, University of Michigan, Ann Arbor, MI 48109, USA \\}
$^{\rm 3}${\it Department of Physics, University of Florida, Gainesville, FL 32611, USA \\}
$^{\rm 4}${\it Institute for Fundamental Theory, University of Florida, Gainesville, FL 32611, USA}

\text

\large

\section*{Abstract}
Cuspy triaxial potentials admit a large number of chaotic orbits, which moreover exhibit extreme ``stickiness'' that makes the process of chaotic mixing surprisingly inefficient. Environmental effects, modeled as noise and/or periodic driving, help accelerate phase space transport but probably not as much as in simpler potentials. This could mean that cuspy triaxial ellipticals cannot exist as time-independent systems.

\section{Introduction}
An enterprise of considerable interest in galactic dynamics is the construction of equilibrium models of galaxies, in the form of time-independent solutions to the collisionless Boltzmann equation. Before the mid-70s observations were consistent with elliptical galaxies possessing spherical or spheroidal (axial) symmetries. However, spherical systems are all integrable, and it is possible to construct realistic axisymmetric systems that contain large measures of regular orbits. For this reason, chaos and the ensuing complications were usually seen more as a curiosity rather than a necessity, and often conveniently ignored.

This situation started changing in the mid-70s, when new kinematical data indicated that many elliptical galaxies rotate too slowly to be rotationally supported. This led Binney (1978) to propose that at least some ellipticals could perhaps be genuinely triaxial objects. In the 80s, improving CCD detectors enabled the acquisition of high quality surface brightness profiles. These revealed a rich structure in the shapes of the isophote curves: there were twisted isophotes, ellipticity gradients, disky and boxy isophotes. Clearly, many elliptical galaxies could not be well approximated as spheroids.

However, if one insists on using classical integrals of the motion to construct equilibrium models of triaxial galaxies, then this lack of symmetries in the shapes of galaxies means that the only known triaxial configurations are variants of rotating Riemann ellipsoids or St\"ackel potentials. These systems are completely integrable, but they are also highly unrealistic: they are not dense enough at the center, their densities fall off too fast at larger radii, and their projected isodensity contours do not exhibit the structure observed in real galaxies. Nevertheless, it was hoped that these models would be adequate approximations of real galaxies in the dynamically important central regions.

These hopes were shattered by high-resolution observations from the ground but mainly from the Hubble Space Telescope over the past 7-8 years, which have shown that most ellipticals do not have soft, extended cores of near-constant density, as required by St\"ackel potentials, but their surface brightness profiles exhibit a power-law cusp towards the center which, furthermore, is often dominated by a supermassive black hole. It turns out that these central singularities create resonances that tend to destabilize the centrophilic box orbits which support the triaxial shape, forcing them to become chaotic. This is perhaps the first time in galactic dynamics where global chaos, far from being something exotic and improbable, may play an important role in the structure and evolution of galaxies.

One rather striking consequence of global stochasticity could be that the very notion of self-consistent equilibria may have to be reconsidered, or even abandoned, at least for certain classes of galaxies. It is generally believed that, after galaxy formation and an ensuing ``violent relaxation'' phase, isolated galaxies settle to, or very close to some equilibrium state, perhaps undergoing small oscillations about~it.

However, this picture may not be entirely accurate. It is known that, if a distribution function  depends only on one isolating integral, then the system \emph{has} to be spherical (Perez \& Aly 1996). Since generic triaxial systems only possess one isolating integral, namely the energy $E$ or the Jacobi integral $E_J$, if they are to exist as equilibrium configurations they would also have to obey so-called \emph{local} integrals of the motion. These correspond to regular orbits which, at least in principle, are perfectly valid building blocks out of which equilibria can be constructed. However, there are some potential problems with this picture. First, it is not known \emph{a priori} whether there are enough regular orbits and of the right shape to support the desired three-dimensional model. Second, even if there are enough suitable orbits, it is not known whether nature (real galaxies) can ``find'' them fast enough (\ie within a Hubble time), especially if they have complex topologies. Third, if the potential admits a large measure of chaotic orbits, they will, in general, slowly diffuse through cantori or Arnold webs, and may cause non-negligible secular evolution over a Hubble time. Thus, a galaxy may look nearly in equilibrium in terms of its bulk properties, but be distinctly out of equilibrium from the standpoint of dynamics.

\section{Phase space transport in the triaxial Dehnen potential}
The present work (\cf Siopis \& Kandrup 2000) involved an investigation of some of the dynamical consequences of combining triaxial symmetry with a central density cusp. This was done using a somewhat realistic potential, namely the triaxial generalization of the spherical Dehnen potential, which corresponds to a density:
  \[ \rho(m) = \frac{(3-\gamma)M}{4\pi a b c} \, m^{-\gamma} \,
     (1+m)^{-(4-\gamma)}, \qquad 0 \le \gamma \le 3, \]
  where
  \[ m^2 = \frac{x^2}{a^2} + \frac{y^2}{b^2} + \frac{z^2}{c^2}, \qquad a \ge
  b \ge c \ge 0. \]
The analysis made extensive use of experience and techniques developed earlier for simpler two- and three-dimensional toy potentials. The choice of the triaxial Dehnen potential was made primarily for the purpose of comparing results from this work with published results from other workers (\eg Merritt \& Fridman 1996, Merritt 1997, Wachlin \& Ferraz-Mello 1998). However, it should be mentioned that there are a number of theoretical and numerical shortcomings with the triaxial Dehnen potential: forces are expensive to evaluate numerically, since the potential cannot be written analytically, and it does not reproduce certain features of observed ellipticals, such as twisted isophotes and a variation of $\gamma$ with radius. Here are some of the questions which were addressed in this work:

{\bf 1. What fraction of the accessible phase space is chaotic and what fraction is regular? Do there exist the requisite orbit families for the skeleton of a self-consistent model?} It was found that, in general, the fraction of strongly chaotic orbits increases with $\gamma$ as well as with decreasing $m$ (distance from the center). It ranges from 30\% at the value of $m$ that encloses 5\% of the mass for the $\gamma=2$ model down to 15\% at the half-mass $m$ of a $\gamma=1$ model. It should be emphasized that these are only \emph{lower limits:} many of the orbits classified as ``regular'' may actually be ``sticky'' chaotic. Furthermore, the choice of the initial conditions had the unfortunate side-effect (in order to facilitate comparison with results from other workers) of favoring regular orbits.

{\bf 2. How unstable are individual chaotic orbit segments, i.e., how large are the maximal (short time) Lyapunov exponents $\chi$?} There are two principal conclusions here. First, chaotic orbits tend to be extremely sticky, even for integration times of $20,000 \, t_D$ or longer. This result was also confirmed using a toy potential corresponding to a superposition of an anisotropic oscillator and a Plummer sphere (Kandrup \& Sideris 2002), which strongly suggests that \emph{extreme stickiness may be generic for all cuspy potentials.} Second, even though the (maximal) dimensional Lyapunov exponent (in units of physical time) varies enormously for different choices of $\gamma$ and energy $E$, the dimensionless Lyapunov exponent (in units of dynamical time) does not. This means that, although physically the mechanisms that cause the instability remain the same, chaos should be much more important at low energies and/or for steep cusps and/or when there is a central black hole.

{\bf 3. How efficient is chaotic mixing? Does it make sense to use ``sticky'' chaotic orbits as near-regular building blocks in galaxy models?} In agreement with what was found in simpler toy potentials, an initially localized ensemble of orbits exhibits a \emph{two-stage evolution}. First is observed a rapid approach towards a near-uniform population of the easily accessible regions of phase space, which proceeds on time scales $1/\chi$. This near-equilibrium stage is followed by a much slower evolution, as orbits diffuse along the Arnold web to probe the entire accessible phase space. However, phase space transport is so inefficient, due to the aforementioned extreme stickiness, that (a) the rate of  mixing for a given ensemble can depend on the phase space direction being probed, (b) ensembles with different initial conditions can reach near-equilibrium at very different rates, and therefore (c) one should not expect that the near-equilibrium should be independent of the choice of chaotic ensemble. Therefore, one should be very careful when using sticky orbits as time-independent building blocks, \eg when calculating a library of orbits for Schwarzschild's method. One should make sure that the orbits have been integrated long enough to probe the true invariant measure --which is hard to do when stickiness persists even after tens of thousands of dynamical times-- or realize that the constructed model will not be a true equilibrium, but will evolve in time.

{\bf 4. What is the impact of internal and external irregularities, such as internal graininess, or the presence of companion galaxies, or the presence of other galaxies in a dense cluster?} We can probe this question by modifying the force as follows:
\[ \mathbf{F} \quad = \quad -\nabla\Phi_{\mbox{\footnotesize Dehnen}} \quad + \quad \sum_{i} \mathbf{A}_i \, \sin(\mathbf{\omega}_i t + \mathbf{\phi}_i) \quad - \quad \left( \eta\mathbf{v} + \mathbf{F}_{\mbox{\footnotesize noise}} \right)
\]
Now the force (per unit mass) $\mathbf{F}$ is determined not only by the bulk potential $\nabla\Phi_{\mbox{\footnotesize Dehnen}}$, but also by a superposition of periodic driving components with different amplitudes $\mathbf{A}_i$, frequencies $\mathbf{\omega}_i$ and phases $\mathbf{\phi_i}$ (emulating the presence of companion galaxies) as well as by the presence  of dynamical friction $\eta\mathbf{v}$ and noise $\mathbf{F}_{\mbox{\footnotesize noise}}$. Two different kinds of noise were employed: additive white noise and colored noise of the Ornstein-Uhlenbeck type (Pogorelov \& Kandrup 1999). The white noise (zero autocorrelation time) is intended to mimic the discreteness effects in the interior of a galaxy (Chandrasekhar 1943) whereas colored noise (finite autocorrelation time) is intended to mimic the effects of an external stochastic environment where the duration of interactions can be comparable to the dynamical times inside the galaxy. It was found, again in accordance with earlier results in simple 2-D and 3-D potentials, that even small amounts of white noise can dramatically accelerate the rate of phase space transport. Colored noise and/or periodic driving can have a similar effect provided their power spectra contain enough power at frequencies comparable to the natural frequencies of the orbits, which suggests strongly that this is a resonance phenomenon. A corollary of this finding is that the libraries of orbits for Schwarzschild models should be constructed using noisy integrations of the orbits. In this way, many sticky orbits will become unstuck, thus yielding a better approximation of the sought-after invariant measure.

{\bf 5. How are these results affected by the presence of a central supermassive black hole?} It was found that a central supermassive black hole tends to increase the fraction of strongly chaotic orbits, thus enhancing the strength of the central cusp. Not surprisingly, the effect of a black hole is stronger for weaker cusps. However, when measured in units of dynamical time, a central black hole does \emph{not} reduce the observed level of stickiness, and thus it does \emph{not} accelerate the convergence to a true invariant distribution. However, in terms of physical time the presence of a black hole may accelerate phase space transport as it reduces the dynamical time of the system.

It seems that the cause for the increase of the fraction of chaotic orbits in the presence of a central black hole is the appearance of resonance overlaps due to the superposition of two different symmetries, namely the spherical symmetry of the black hole and the triaxial symmetry of the Dehnen potential. However, this may also indicate that the increase of chaotic orbits could actually be unrealistic and exaggerated: many elliptical galaxies gradually become more axisymmetric towards the center, and this may not be accidental.

\section{Implications for cuspy triaxial galaxies}
The research reported in this paper is being extended in three waysby considering (i) variations of the axis ratio, (ii) different types of colored noise, and (iii) other, more realistic triaxial potentials. However, it seems that a new picture is emerging as to whether cuspy triaxial elliptical galaxies can attain time-independent triaxial
configurations. The crucial element in this picture is the extreme stickiness manifested by orbits in the triaxial Dehnen potential, which can dramatically reduce the efficiency of chaotic mixing on a global scale. If this is a generic characteristic of cuspy triaxial potentials, it would mean that, within a Hubble time, triaxial ellipticals may not be able to completely phase mix away irregularities that were either inherited from their formation stage or were induced in the course of later gravitational interactions with other galaxies. This, in turn, would imply that they could be undergoing continuous evolution, which could be either secular (such as evolution towards a more axisymmetric state, or continuous transformation to a succession of nonaxisymmetric shapes) or semi-periodic (such as ``wobbling'' about some near-equilibrium state).

It is also interesting to note that noise and periodic driving with physically plausible characteristics help accelerate phase space transport in the triaxial Dehnen potential, as they did in simpler toy potentials. However, although in the simpler potentials they would suffice to make chaotic mixing an efficient process within a Hubble time, it is not clear whether this is still the case with cuspy triaxial potentials, where stickiness is so pervasive and persistent. Careful construction of Schwarzschild models may help answer some of these questions.

\section*{References}
Binney J., 1978, Comments on Astrophysics, 8, 27 \\
Chandrasekhar S., 1943, Rev.\ Mod.\ Phys., 15, 1 \\
Kandrup H.\ E., Sideris I.\ V., 2002, Celestial Mech., in press [astro-ph/0111517] \\
Merritt D., 1997, ApJ, 486, 102 \\
Merritt D., Fridman T., 1996, ApJ, 460, 136 \\
Perez J., Aly J.-J., 1996, MNRAS, 280, 689 \\
Pogorelov I.\ V., Kandrup H.\ E., 1999, Phys.\ Rev., E60, 1567 \\
Siopis C., Kandrup H.\ E., 2000, MNRAS, 319, 43 \\
Wachlin F.\ C., Ferraz-Mello S., 1998, MNRAS, 298, 22 \\
\end{document}